\documentclass[reprint, aps,prl,groupeaddress, longbibliography]{revtex4-2}

\usepackage{graphicx}	
\usepackage{dcolumn}	
\usepackage{bm}		
\usepackage{SIunits}	
\usepackage{xcolor}
\usepackage{booktabs}
\usepackage{float}
\usepackage[normalem]{ulem}
\usepackage{hyperref}

\definecolor{r}{rgb}{0.86,0.08,0.23}
\definecolor{blue_n}{rgb}{0.,0.3,0.5}

\hypersetup{
backref=true, 
pagebackref=true,
hyperindex=true, 
colorlinks=true, 
breaklinks=true, 
citecolor = blue_n, 
urlcolor= blue_n,
linkcolor= black, 
pdftitle={Single G centers in silicon fabricated by co-implantation with carbon and proton}, 
pdfauthor={Yoann~Baron,  Alrik~Durand, Tobias~Herzig, Mario~Khoury, S\'ebastien~Pezzagna, Jan~Meijer, Isabelle~Robert-Philip, Marco~Abbarchi, Jean-Michel Hartmann, Shay Reboh, Jean-Michel~G\'erard, Vincent~Jacques, Guillaume~Cassabois and Ana\"is~Dr\'eau } 
}

\begin{document}

\title{Detection of single W-centers in silicon}

\author{Yoann~Baron$^{1}$}	 
\thanks{These authors contributed equally to this work.}	
\author{Alrik~Durand$^{1,\ast}$} 		
\author{P\'eter~Udvarhelyi$^{2}$}
\author{Tobias~Herzig$^3$}
\author{Mario~Khoury$^4$}			
\author{S\'ebastien~Pezzagna$^3$}		
\author{Jan~Meijer$^3$}		
\author{Isabelle~Robert-Philip$^1$}	
\author{Marco~Abbarchi$^4$}
\author{Jean-Michel Hartmann$^5$}
\author{Vincent Mazzocchi$^5$}		
\author{Jean-Michel~G\'erard$^6$}
\author{Adam~Gali$^{2,7}$}		
\author{Vincent~Jacques$^1$}		
\author{Guillaume~Cassabois$^1$}		
\author{Ana\"is~Dr\'eau$^1$}		
\email{anais.dreau@umontpellier.fr}	

\affiliation{$^1$Laboratoire Charles Coulomb, Universit\'e de Montpellier and CNRS, 34095 Montpellier, France} 
\affiliation{$^2$Wigner Research Centre for Physics, P.O. Box 49, H-1525 Budapest, Hungary}
\affiliation{$^3$Division of Applied Quantum Systems, Felix Bloch Institute for Solid State Physics, University Leipzig, Linn\'estra\ss e 5, 04103 Leipzig, Germany}
\affiliation{$^4$CNRS, Aix-Marseille Universit\'e, Centrale Marseille, IM2NP, UMR 7334, Campus de St. J\'er\^ome, 13397 Marseille, France }
\affiliation{$^5$Univ. Grenoble Alpes and CEA, LETI,  F-38000 Grenoble, France} 
\affiliation{$^6$Univ. Grenoble Alpes, CEA, Grenoble INP, IRIG-PHELIQS, F-38000 Grenoble, France}
\affiliation{$^7$Department of Atomic Physics, Budapest University of Technology and Economics, Budafoki \'ut 8., H-1111 Budapest, Hungary}

\begin{abstract}
Controlling the quantum properties of individual fluorescent defects in silicon is a key challenge towards advanced quantum photonic devices prone to scalability. 
Research efforts have so far focused on extrinsic defects based on impurities incorporated inside the silicon lattice. 
Here we demonstrate the detection of single intrinsic defects in silicon, which are linked to a tri-interstitial complex called W-center, with a zero-phonon line at 1.218$\mu$m. 
Investigating their single-photon emission properties reveals new information about this common radiation damage center, such as its dipolar orientation and its photophysics. 
We also identify its microscopic structure and show that although this defect does not feature electronic states in the bandgap, Coulomb interactions lead to excitonic radiative recombination below the silicon bandgap. 
These results could set the stage for numerous quantum perspectives based on intrinsic luminescent defects in silicon, such as integrated quantum photonics and quantum communications.

\end{abstract}
\maketitle


\def\thefootnote{*}\footnotetext{These authors contributed equally to this work}\def\thefootnote{\arabic{footnote}}

\section*{Introduction}

The convergence between material science and quantum technologies is a current trend reviving the field of fluorescent point defects in semiconductors~\cite{zhang_material_2020}. 
A prominent example is the successful application of nitrogen-vacancy defects in diamond~\cite{doherty_nitrogen-vacancy_2013} to quantum information science~\cite{awschalom_quantum_2018} and quantum sensing~\cite{barry_sensitivity_2020}.
These works have stimulated a quest to isolate other individual solid-state artificial atoms in a wide range of semiconductors, including silicon carbide~\cite{castelletto_silicon_2020}, hBN~\cite{zhang_point_2020} and, more recently, silicon~\cite{redjem_single_2020,durand_broad_2021}. 
In a reverse perspective, the ability to investigate these color centers down to the ultimate level of single defects can boost our understanding of the physics of these complex solid-state systems.
This new asset could be especially pivotal for silicon, for which the control of defect engineering is essential for numerous applications~\cite{yoshida_defects_2015}, primarily with microelectronics~\cite{fleetwood_defects_2008}. 
Applying the quantum optics toolbox to single defects in silicon could allow to tackle long-standing issues, such as the defect microscopic structure, still unresolved even for some of the most common color centers~\cite{pichler_intrinsic_2004}. 
Here we illustrate this approach with the so-called W-center in silicon~\cite{davies_1018_1987, davies_optical_1989}, an intrinsic defect frequently observed after radiation damage~\cite{davies_radiation_2006, chartrand_highly_2018}.

\begin{figure}[h!]
  \includegraphics[width=0.95\columnwidth]{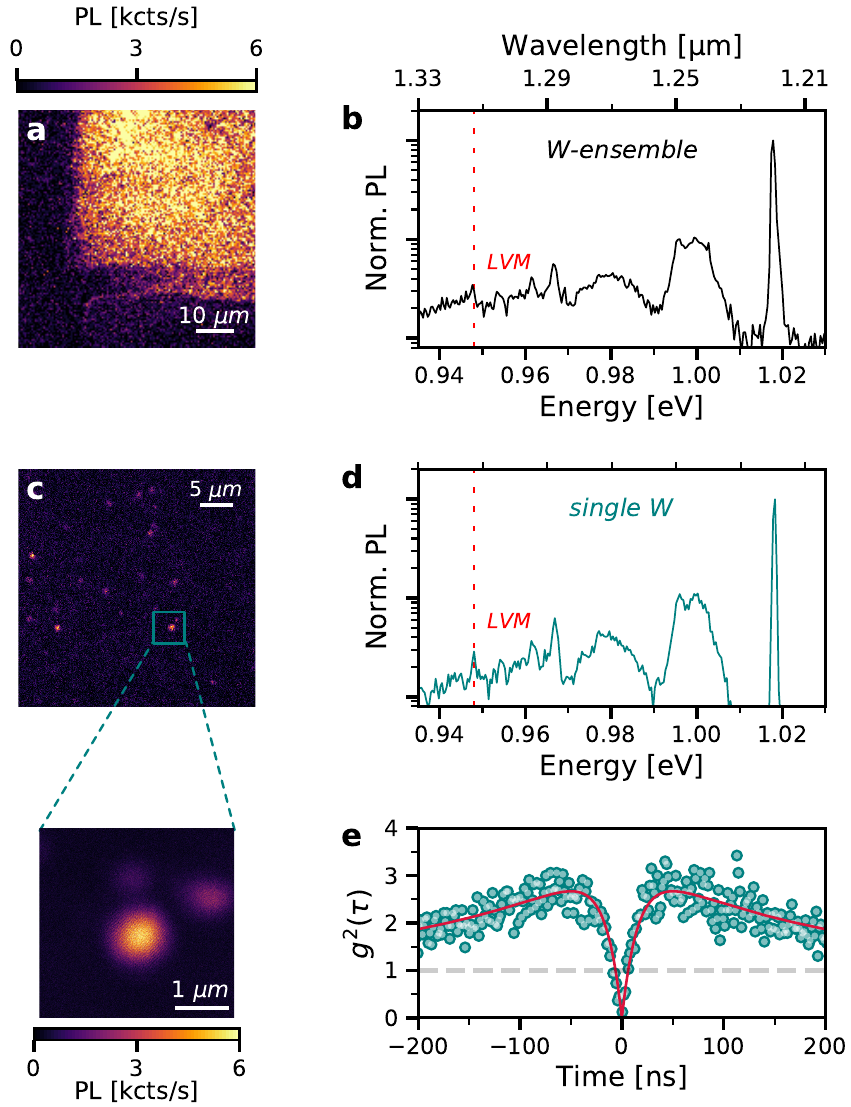}
  \caption{Isolation of single W-centers in silicon. 
  \textbf{a.} PL raster scan of an area of the SOI sample that has been locally implanted with Si ions. 
  \textbf{b.} PL spectrum recorded on a position inside the implanted area, revealing the formation of an ensemble of W-centers. 
  The dashed red line indicates the energy of the LVM at 70 meV \cite{davies_1018_1987, hayama_photoluminescence_2004}. 
  \textbf{c.} PL raster scan recorded away from the highly-implanted region. 
 \textbf{d.} PL spectrum from the bright spot in \textbf{c}, square zone with zoom. 
 \textbf{e.} Second-order autocorrelation function g$^2$($\tau$) measured onto this emitter. 
  The antibunching effect at zero delay reaches the value $g^{(2)}(0)\simeq 0.12$, without any background or noise correction.
  The red curve is data fitting with the 3-level model used in \cite{redjem_single_2020}. 
  All measurements are acquired at 10 K. 
 }
  \label{fig:first}
\end{figure}

The W-center in silicon, characterized by an intense zero-phonon line emission at 1.018 eV (1218 nm), has attracted a lot of attention in the literature~\cite{pichler_intrinsic_2004}.
Firstly, it is made of self-interstitial atoms, that limit the performances of silicon nanoelectronic devices when unintentionally produced during nanofabrication~\cite{stolk_physical_1997}. 
Secondly, its presence is correlated with the degradation of silicon-based components in environments subject to strong radiations~\cite{davies_radiation_2006} like in particle physics experiments~\cite{macevoy_defect_1997}. 
This intrinsic defect can also be created by silicon implantation~\cite{buckley_optimization_2020}, neutron irradiation~\cite{surma_new_2008} or following laser annealing~\cite{skolnick_defect_1981}, and in isotopically purified $^{28}$Si which enables to drastically reduce its optical linewidth~\cite{chartrand_highly_2018}.
Recently, the W-center has seen renewed interest for Si-based classical photonics, where it has been used as an active medium in LEDs~\cite{bao_point_2007,buckley_all-silicon_2017} and optical resonators~\cite{tait_microring_2020}. 
At last, while it is widely admitted that the W-center in silicon relates to a small cluster of interstitials~\cite{giri_evidence_2001}, the identification of the atomic structure of the W-defect is still controversial~\cite{pichler_intrinsic_2004,coomer_interstitial_1999, gharaibeh_molecular-dynamics_1999, jones_self-interstitial_2002, richie_complexity_2004, lopez_structure_2004, carvalho_density-functional_2005, santos_insights_2016}.

In this paper, we demonstrate the first optical isolation of individual W-centers in silicon and close the long-lasting debate about its nature and atomistic configuration. 
Our optical characterization and quantum optics experiments at the single-defect level are confronted with state-of-the-art first-principles simulations to identify the structure of this tri-interstitial complex. 
At last, we show how it is optically active despite the absence of energy levels in the bandgap of the host semiconductor.


\section*{Experimental results}
We study W-centers formed by $^{28}$Si implantation in a silicon-on-insulator (SOI) wafer at 65 keV followed by a flash annealing for recrystallization (see Methods for details).  
Localized implantations have been carried out to create high defect concentration areas and low defect concentration at their periphery. 
These low-density areas enable to isolate single W-centers and to analyze their properties by micro-photoluminescence spectroscopy.

\noindent
\textbf{Detection of single W-centers.}
Optical scans of the sample performed at 10 K under 532-nm laser excitation show a high PL intensity at the location of the implanted areas (Fig. \ref{fig:first}\textbf{a}).
As shown in Figure~\ref{fig:first}\textbf{b}, the corresponding emission is typical of the W-center in silicon, with a zero-phonon line (ZPL) at 1.018 eV (1218 nm) and a broad structured phonon-sideband that includes an emission line associated with a local vibrational mode (LVM) at 70 meV \cite{davies_1018_1987, hayama_photoluminescence_2004}.
Few hundreds of microns away from the highly-implanted zones, optical scans reveal isolated PL hotspots (Fig. \ref{fig:first}\textbf{c}) that show exactly the same PL spectrum (Fig. \ref{fig:first}\textbf{d}) and are thus also attributed to the emission of one (or several) W-centers. 
The Debye-Waller factor, that expresses the proportion of photons emitted into the ZPL, reaches $\simeq 40$\% in these hotspots (see SI), as in ensemble measurements~\cite{davies_1018_1987}. 
In order to check whether they stem from single defects, we use the quantum optics technique of intensity-correlation. 
The second-order autocorrelation function measured with a Hanbury-Brown and Twiss interferometer on such an isolated PL spot features a strong antibunching at zero delay, with $g^{(2)}(0) = 0.12 \pm 0.05$ (Fig. \ref{fig:first}(c)). 
This value, not corrected from any background counts or noise, is well below the single-photon emission threshold of 0.5 \cite{beveratos_room_2002}, thus evidencing the presence of a single emitter and therefore of an individual W-center. 
Furthermore, a bunching effect corresponding to $g^{(2)}(\tau)>1$ is revealed in the $g^{(2)}(\tau)$ plot. 
It indicates that the relaxation involves a non-radiative path through a metastable level~\cite{beveratos_room_2002}, an information beyond the reach of ensemble measurements.\\

\begin{figure}[h!]
  \includegraphics[width=\columnwidth]{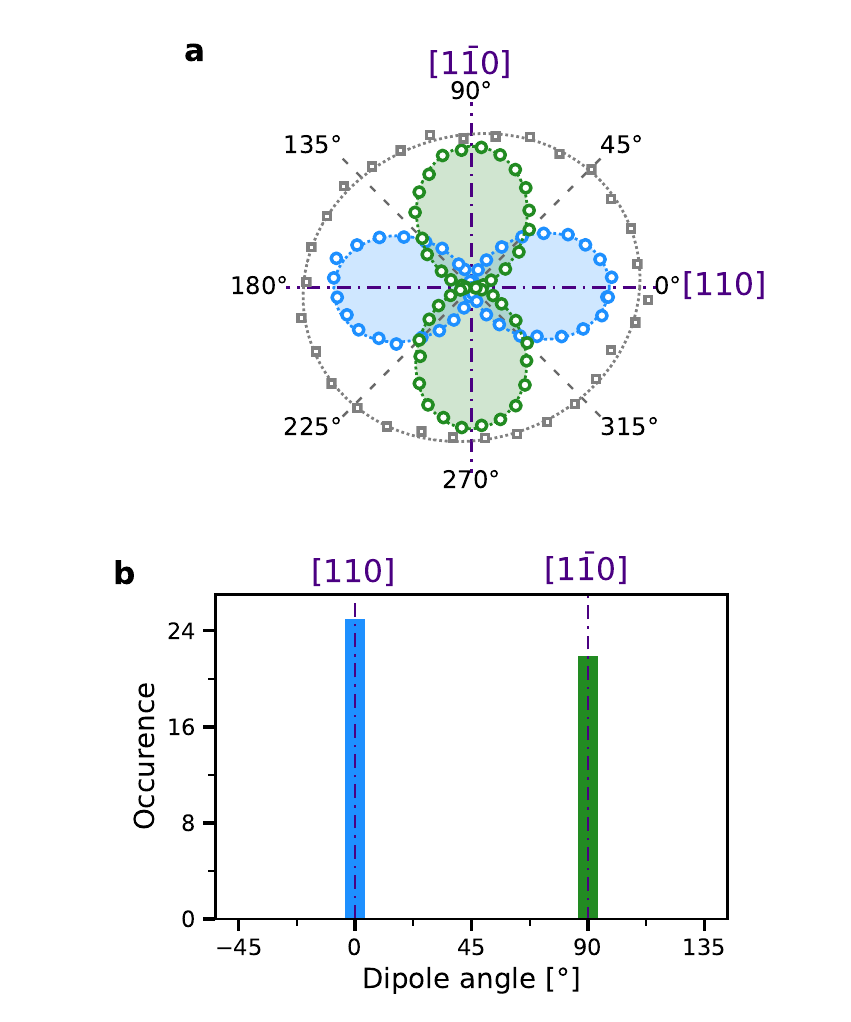}
  \caption{Polarization of the single photons emitted by W-centers.
  \textbf{a.} Emission polarization diagrams recorded on 2 individual W-defects (circle markers). 
 The PL signal is measured for different angular positions of a half-waveplate rotated in front of a polarizer in the detection path \cite{lethiec_measurement_2014}. 
   The 0$\degree$ and 90$\degree$ directions match the crystal axes [110] and [1$\bar{1}$0].
   Solid lines are fits using a cos$^2$($\theta$) function.
 The emission polarization diagram recorded on the ensemble of W-centers is shown in comparison (square markers). 
 Its slightly oblong shape is attributed to polarization distortion induced by the optical setup. 
   \textbf{b.} Histogram of the emission dipole angle $\theta$ measured on a set of 47 individual W-centers.}
  \label{fig:polar}
\end{figure}

\noindent
\textbf{Orientation of the emission dipole of the W-centers.}
The polarization analysis of the single photons emitted by these defects provides information about the orientation of their emission dipoles \cite{lethiec_measurement_2014}. 
Whereas the PL emission is unpolarized in W-centers ensembles, a strong linear polarization is observed at the single defect scale, as shown in Figure \ref{fig:polar}\textbf{a}. 
The emission polarization diagrams recorded on two typical W-centers display a visibility exceeding $96\%$, indicating the emission of linearly polarized photons~\cite{lethiec_measurement_2014}, and thus accounting for the presence of a single emitting dipole.
A statistical analysis over a set of 47 individual W-centers demonstrates that this dipole projected onto the $(001)$ sample surface can take only two possible orientations with equal probability: either along $[110]$ or $[1\bar{1}0]$~(Fig. \ref{fig:polar}\textbf{b}). 
The existence of a single emission dipole for the W-center in silicon is a key information that will be corroborated later by the advanced first-principles calculations to confirm the identification of the defect structure.\\

\begin{figure*}[ht]
  \includegraphics[width=0.7\textwidth]{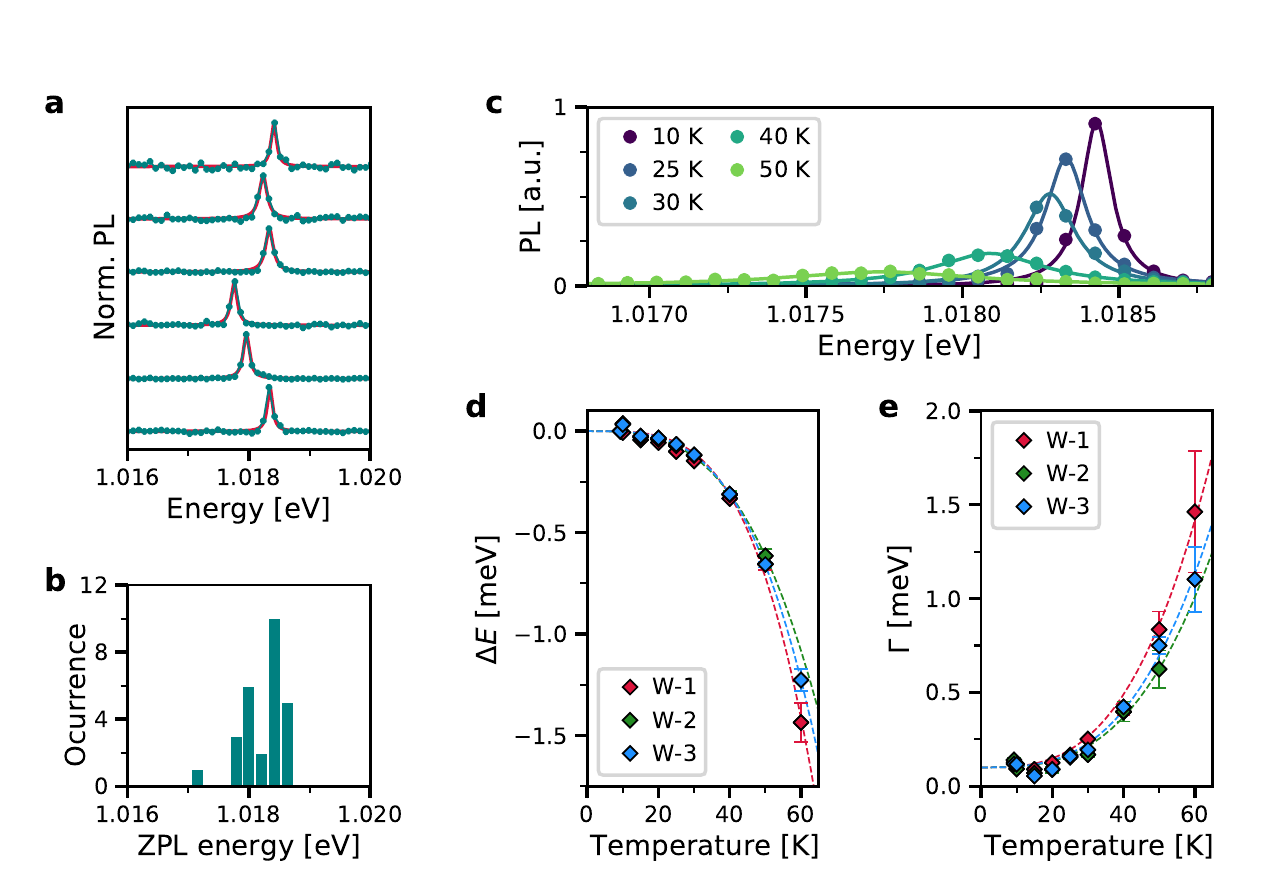}
  \caption{Statistical analysis and evolution with temperature of the ZPL of single W-centers.
   \textbf{a.} ZPL spectra recorded on individual W-defects.
   The red solid lines are data fitting with a Lorentzian function to extract the ZPL energy position. 
   The FWHM is limited by the spectrometer resolution to {$\simeq$~0.10} meV. 
   \textbf{b.} Distribution of the ZPL energy for a set of 27 single W-centers.
   \textbf{c.} ZPL spectra recorded on a single W-center at different temperatures. 
   The lines are fitted with a Lorentzian function (solid curves). 
   \textbf{d.} Evolution versus temperature of the ZPL energy shift $\Delta E$ with respect to the ZPL energy at 8K for three single W-defects.
   \textbf{e.} Evolution with temperature of the FWHM $\Gamma$ for the same defects. 
   Error bars represent the standard error from data fitting results.
   Dashed lines in \textbf{d}-\textbf{e} are data fitting results with a polynomial function (see SI for details). 
 }
  \label{fig:zpl_lifetime}
\end{figure*}

\noindent
\textbf{ZPL statistics and evolution with temperature.}
Single-defect detection further enables to investigate statistical variations in the properties of W-centers in silicon, such as the ZPL energy fluctuations. 
As visible on Figures {\ref{fig:zpl_lifetime}\textbf{a}-\textbf{b}}, ZPL energy variations within 1 meV are detected from one W-center to another. 
This effect is attributed to different strain and electrostatic environment experienced by each emitter \cite{lindner_strongly_2018}. 
We note that these fluctuations are about 20 and 6 times smaller than those observed in 220-nm thick silicon layer on single G-centers~\cite{redjem_single_2020} and the unidentified color centers labelled as SD-2 in Ref.~\cite{durand_broad_2021}, respectively. 
This observation is consistent with the small sensitivity to stress variations previously reported for ensembles of W-centers in silicon \cite{davies_1018_1987}.

The evolution of the ZPL with temperature is then analyzed for individual W-centers.~Two effects are observed when increasing the temperature: a red-shift of the zero-phonon line along with its broadening (Fig. \ref{fig:zpl_lifetime}\textbf{c}). 
The increase of the ZPL width is due to phonon-assisted broadening processes, whereas the temperature-dependence of the electron-phonon interaction is thought to be the dominant contribution to the ZPL red-shift for the W-center in silicon~\cite{davies_1018_1987}. 
As shown in Figures \ref{fig:zpl_lifetime}\textbf{d}-\textbf{e}, the temperature evolutions of both the ZPL shift and FWHM for individual W-centers are identical within error bars, and consistent with former W-ensemble measurements~\cite{davies_1018_1987}.
These two parameters follow a polynomial law with temperature (see SI), which is commonly observed for fluorescent defects in semiconductors~\cite{jahnke_electronphonon_2015}. \\

\begin{figure}[h!]
  \includegraphics[width=\columnwidth]{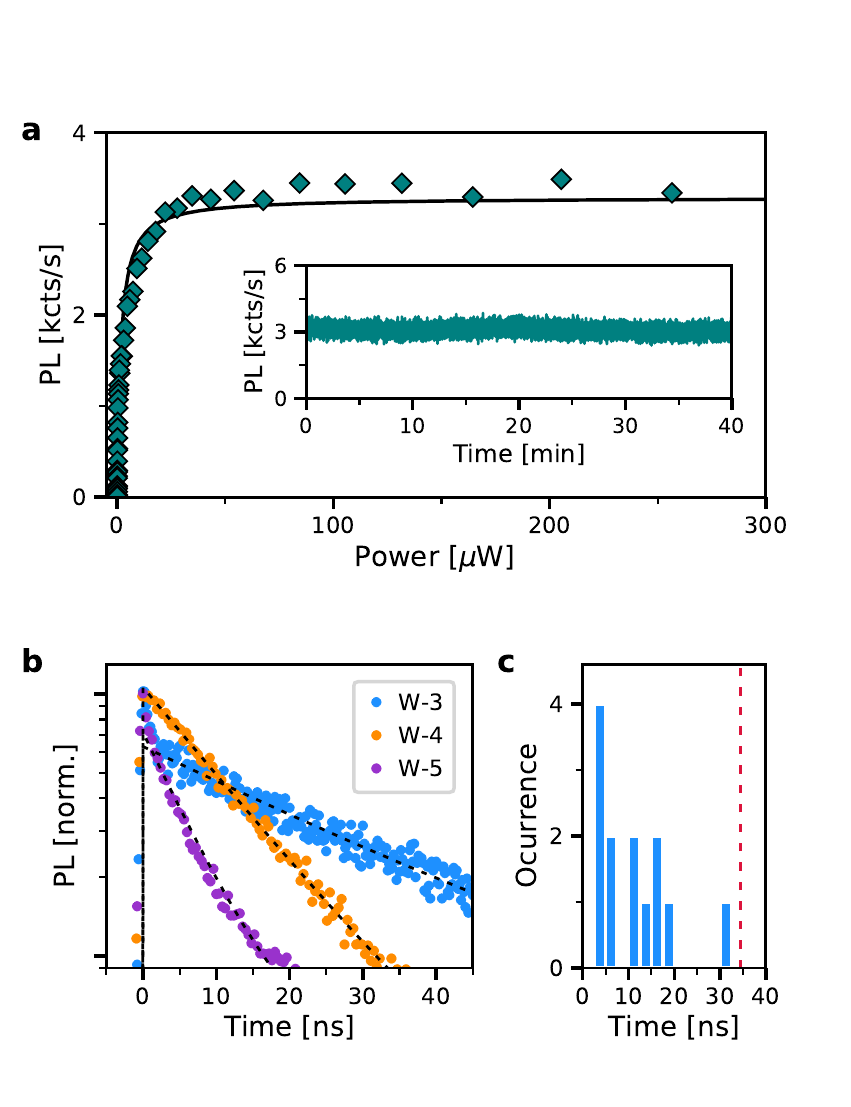}
  \caption{Photophysics of single W-centers.
  \textbf{a.} Standard PL saturation curve recorded on an individual W-defect. 
  The solid line is data fitting with the usual saturation function: $I_{sat}/(1+P_{sat}/P)$ leading to a saturation power $P_{sat}\simeq 2\,\mu$W (see main text). 
  Inset: Typical PL time-trace measured at 30 $\mu$W on a single defect. 
    \textbf{b.} Time-resolved PL recorded for 3 W-centers under a 50-ps pulsed-laser excitation at 532 nm.
   The excited state lifetime is extracted from data fitting with a single exponential function (dashed lines), which gives respectively $30.8 \pm 0.3$ ns, $12.7 \pm 0.1$ ns and $7.1 \pm 0.1$ ns for W-3, W-4 and W-5.
   The sharp peak at the beginning of the pulse for W-3 comes from background counts. 
   \textbf{c.} Histogram of the excited-state lifetime measured on a set of 13 W-centers. 
   The red dashed line represents the excited-state lifetime value reported from an ensemble measurement \cite{buckley_optimization_2020}. }
  \label{fig:photophysics}
\end{figure}


\noindent
\textbf{Photophysics of single W-centers.}
The maximum count rates from individual W-centers are estimated by recording their PL evolution with illumination power. 
For a simple 2-level system, the PL intensity increases then saturates above a given power referred to as the saturation power $P_{sat}$. 
Such a saturation curve measured on a single W-center is shown in Figure \ref{fig:photophysics}\textbf{a}.
The intensity at saturation $I_{sat}$ reaches typically 2-6 kcounts/s for individual W-centers.  
These count rates are mainly limited by the photon loss linked to total internal reflection associated with the high refractive index of silicon (n$ \simeq 3.5$) and the weak detection efficiency of our detectors (10\%). 
A statistical analysis over a few tens of W-defects shows that this standard saturation behavior is only observed on roughly one third of the defects. 
The rest of the emitters displays anomalous saturation curves that could result from the coupling to a dark state (see SI for more details). 

The single-photon emission of individual W-defects can be photostable on hour-timescale, as visible on the inset of Figure \ref{fig:photophysics}\textbf{a}. 
Data acquisitions exceeding 50 hours of optical illumination have been performed on single W-centers. 
However, we note that at high optical power, typically above 500 $\mu$W, photobleaching effects may happen for some emitters.

The optical cycle dynamics of individual W-centers is analysed through time-resolved PL measurements under 50-ps laser excitation at 532 nm. 
As shown in Figure \ref{fig:photophysics}\textbf{b}, the PL decreases over time following a mono-exponential decay, but with a characteristic time that strongly changes from one defect to another. 
This measurement, performed on 13 W-defects, indicates that the excited state lifetime can vary by one order of magnitude from roughly 3 ns up to 30 ns (Fig. \ref{fig:photophysics}\textbf{c}). We note that all the measured values are shorter than the excited-level lifetime recently reported on an ensemble of W-centers that reaches $34.5 \pm 0.5$ ns \cite{buckley_optimization_2020}. 
Since our measurements deal with W-centers embedded in a 60-nm thick Si layer (see Methods), the lifetime fluctuations could be explained by non-radiative recombination channels due to the proximity of interfaces.

\begin{figure*}[ht]
  \includegraphics[width=0.7\textwidth]{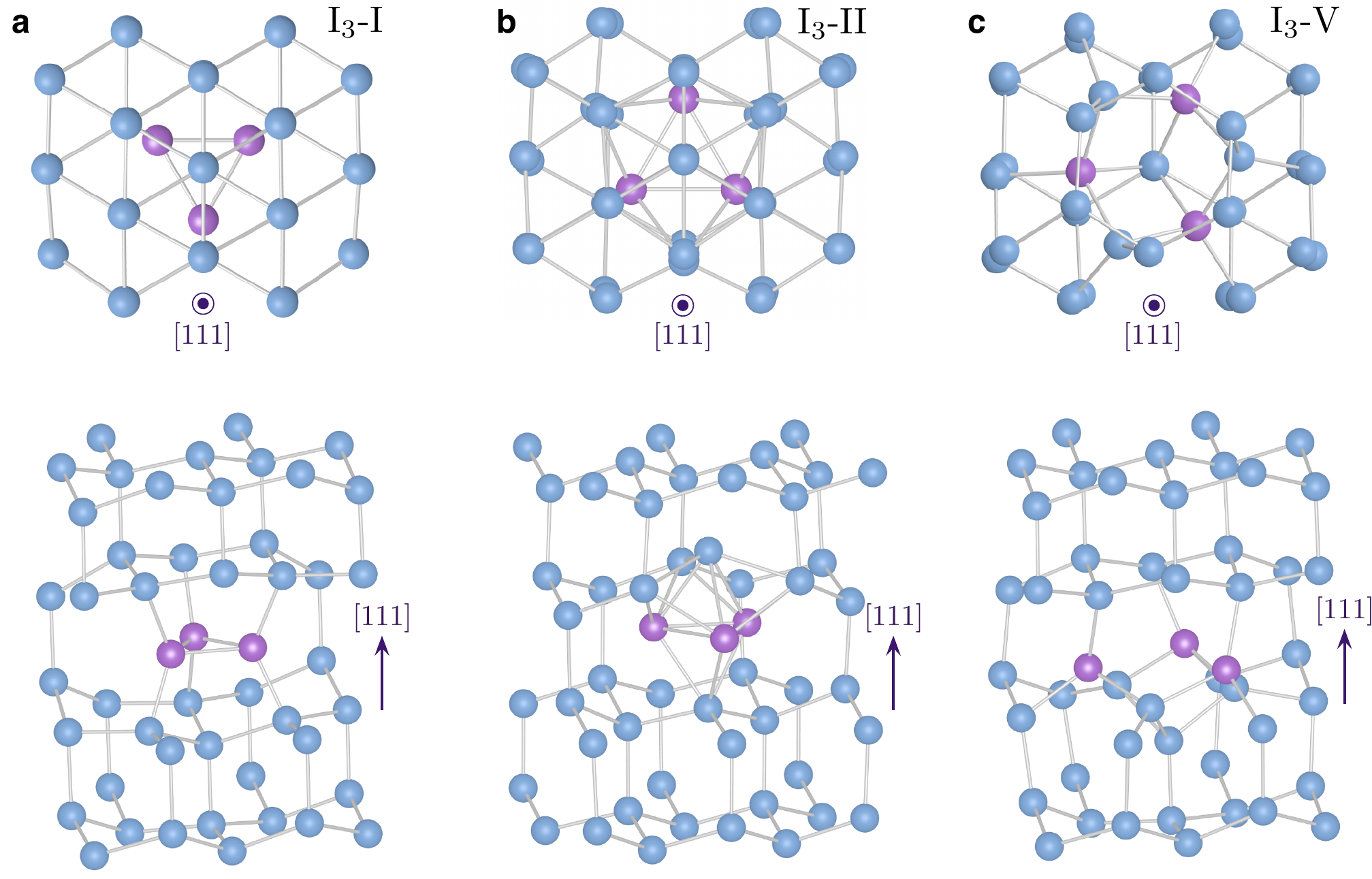}
  \caption{Tri-interstitial configurations investigated as potential candidates for the W-center in silicon. \textbf{a.} I$_3$-I defect, \textbf{b.} I$_3$-II defect and \textbf{c.} I$_3$-V defect.}
  \label{fig:geom}
\end{figure*}

\section*{Identification of the W-center structure}

\noindent
\textbf{Candidates for the W-center in silicon.}
Several models have been proposed to account for the W-center PL, such as $\left<111\right>$-split-triple di-interstitial~\cite{coomer_interstitial_1999}, tri- and tetra-interstitial~\cite{coomer_interstitial_1999,gharaibeh_molecular-dynamics_1999,jones_self-interstitial_2002}. 
These theoretical studies used calculations based on tight-binding models~\cite{sankey_ab_1989} and density functional theory (DFT) within the local density approximation (LDA)~\cite{ceperley_ground_1980}. The tri-interstitial (I$_3$) defect in its neutral charge state was found to be the most stable configuration ~\cite{gharaibeh_molecular-dynamics_1999}.
The first proposed tri-interstitial structure (I$_3$-I) featuring both the appropriate trigonal symmetry ($\mathrm{C}_{3v}$) and LVM energy at 70 meV corresponds to a triangular arrangement bridging three adjacent $\left<111\right>$ bonds (Fig. \ref{fig:geom}\textbf{a})~\cite{coomer_interstitial_1999}.
However, later studies have casted some doubts about this assignment as another tri-interstitial defect has been found with a formation energy lower by $\simeq$ 1 eV \cite{lopez_structure_2004,santos_insights_2016}. 
Still this second structure (I$_3$-II) associated with a compact tetrahedral shape (Fig. \ref{fig:geom}\textbf{b}), cannot account for the W-center PL as it possesses the wrong symmetry ($T_d$) and does not present any LVM~\cite{carvalho_density-functional_2005}.
Additionally, a new extended form of the first structure showing $\text{C}_{3}$ symmetry and the correct LVM energies was revealed by tight-binding molecular dynamics calculations~\cite{richie_complexity_2004}.~This third configuration, labeled I$_3$-V from~\cite{carvalho_density-functional_2005}, is displayed on Figure \ref{fig:geom}\textbf{c}. 
This defect was studied using DFT methods with LDA~\cite{carvalho_density-functional_2005} and generalized gradient approximation using the Perdew-Burk-Ernzerhof functional (PBE)~\cite{santos_insights_2016} revealing formation energy between the two previous configurations. 
However the presence of energy levels inside the bandgap to enable its optical activity is still under debate \cite{carvalho_density-functional_2005, santos_insights_2016}. 
In the light of our novel measurements at the single-defect scale, we present advanced theoretical modeling in order to elucidate the W-center structure.\\

\noindent
\textbf{Formation energies}. 
We first compare the formation energy provided by the DFT calculations using Heyd-Scuseria-Ernzerhof  functional (HSE06)~\cite{krukau_influence_2006} for the three tri-interstitial configurations. 
Our calculated formation energies for the neutral I$_3$-I, I$_3$-II, and I$_3$-V configurations are 8.17~eV, 7.69~eV, and 7.52~eV, respectively.
Contrary to the literature where I$_3$-II was presented as the most energetically favorable structure among these three clusters~\cite{carvalho_density-functional_2005, santos_insights_2016}, our results indicate that the I$_3$-V configuration is the most stable defect configuration. These findings are supported by the bonding properties accurately described by the hybrid functional. We calculate the Si-Si bond lengths of $2.547~\text{\AA}, 2.465~\text{\AA}$ and $2.315~\text{\AA}, 2.296~\text{\AA}, 2.242~\text{\AA}$ in the core of I$_3$-II and I$_3$-V defect configurations, respectively. The Si-Si bond lengths in the I$_3$-V defect are closer to the calculated one in the perfect Si crystal at $2.352~\text{\AA}$. Furthermore, the extension of the defect triangle in the I$_3$-V configuration shows bonding properties more compatible with the bulk than that of the I$_3$-II configuration.\\

\begin{figure*}[ht]
  \includegraphics[width=0.85\textwidth]{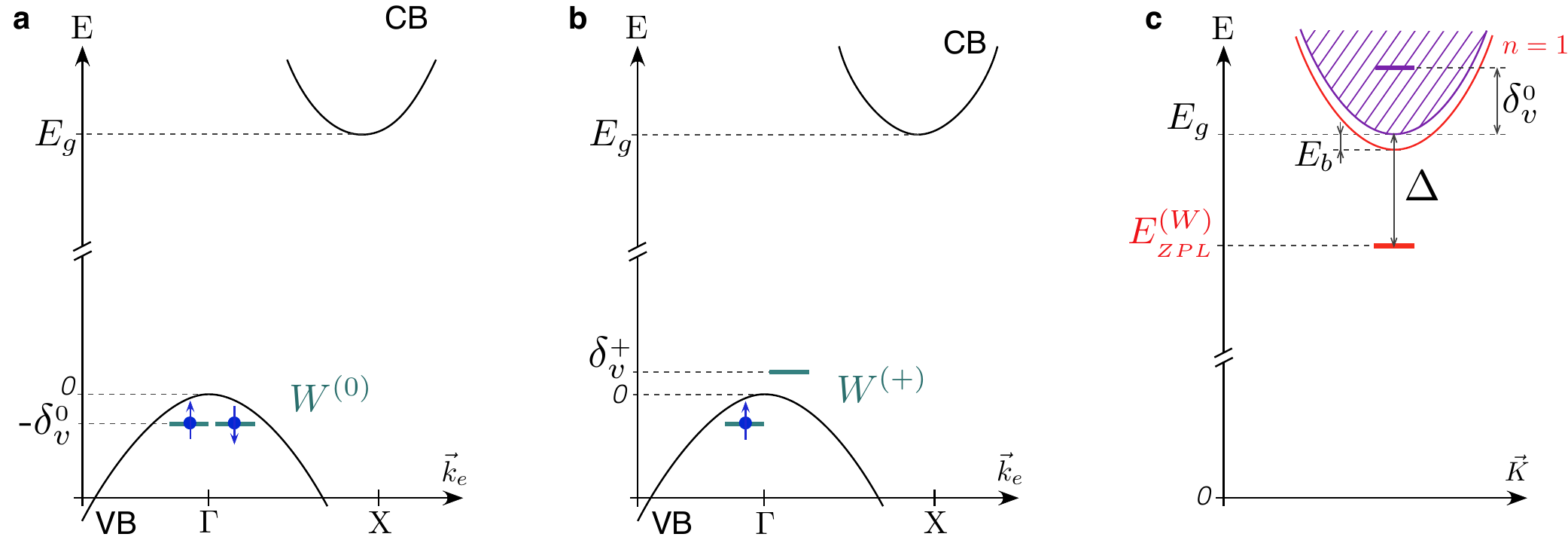}
  \caption{Schematics of the energy level diagrams of the I$_3$-V configuration of the W-center. 
  In the single-particle representation, energy diagrams of \textbf{a.} the ground state of the neutral form of the defect and \textbf{b.} after its ionization.  
  VB: valence band; CB: conduction band; $\vec{k}_e$ is the electron wave vector.
  \textbf{c.} Two-particle representation of the energy levels. 
  The continuum of free electron-hole pairs above the electronic bandgap energy $E_g$ is indicated in purple color. 
  The red parabola represents the ground state of the bulk excitons, whose energy minimum is lower from $E_g$ by the exciton binding energy $E_b$.
  The energy splitting $\Delta$ between the W-center ZPL energy $E_{_{ZPL}}^{_{(W)}}$ and the silicon bandgap includes the energy $\delta_v^+$ and the binding energy of the exciton localized on the defect.
  $\vec{K}$: wave vector of the electron-hole pair center-of-mass. }
  \label{fig:levels}
\end{figure*}

\noindent
\textbf{Local vibrational modes.}
Next, the vibrational properties of the tri-interstitial defects are discussed.
 As the phonon density of states in pristine silicon has a sharp cutoff around 63~meV, the observed 70-meV LVM in the W-center PL spectrum can be used as a fingerprint of the defect structure. 
 Our vibrational calculations show LVMs above this energy cutoff only in the two trigonal defect configurations I$_3$-I and I$_3$-V, and not in the tetrahedral I$_3$-II structure, in agreement with previous studies~\cite{carvalho_density-functional_2005}. 
 In the I$_3$-I configuration, we find 4 strong LVM with inverse participation ratio below 8 (see Methods and SI).
 One of these is the 73.9~meV symmetric deformation mode along $\left<111\right>$ direction localized on the interstitial atoms, corresponding to the only reported LVM by Coomer~{\it et al.}~\cite{coomer_interstitial_1999}. 
 Additionally, we find a double degenerate deformation mode at 74.2~meV and a symmetric stretching mode at 69.9~meV localized on the tri-interstitial atoms. 
 In the I$_3$-V configuration, we find 5 strong LVM with inverse participation ratio below 12. 
Among these, there are 3 totally symmetric modes at 71.6~meV, 65.9~meV and 63.8~meV corresponding respectively to axial stretching of the neighboring Si atoms, twisting of the defect and symmetric breathing (see SI). 
These vibrational modes are slightly shifted to smaller energies compared to the corresponding previously reported calculations in smaller supercells~\cite{carvalho_density-functional_2005, santos_insights_2016}. 
Since both trigonal configurations possess LVM with the appropriate energy around 70 meV, the distinction between I$_3$-I and I$_3$-V cannot be made on the basis of the sole LVM energy \cite{carvalho_density-functional_2005, santos_insights_2016}.\\

\noindent
\textbf{Electronic structure and optical activity.}
We continue by discussing the electronic structure and optical properties of the I$_3$-I and I$_3$-V defect configurations.
Fluorescent defects are usually associated with dangling or elongated bonds featuring weaker binding energies than the bulk lattice bonds. 
As a result, these centers possess energy levels deep inside the semiconductor bandgap. 
On the other hand, defects having LVM with energies higher than the bulk modes exhibit stronger bonding properties due to the increased electrostatic interaction between the defect orbitals.
This feature leads to shallow defect levels in the bandgap, or resonant states close to the band extrema. 
Indeed, no strongly localized in-gap defect levels emerged in our HSE06 DFT calculations for these two tri-interstitial defects in their neutral singlet ground state. 
Although it may contradict an optical emission below the silicon bandgap, we will show that Coulomb correlation effects are prominent in such a defect, a novel feature requiring advanced first-principle calculations for catching the physics of point defects in silicon.

In the I$_3$-V configuration, we find a single $a$ level at $\delta_v^0=73$~meV below the valence band maximum (VBM) (Fig. \ref{fig:levels}\textbf{a}). 
This is a resonant state, that is localized at the defect position in real space while being below the VBM.
After ionization, the unoccupied defect level emerges inside the band-gap resulting in a $0/+$ charge transition level at $\delta_v^+ = 55$~meV above the VBM with Freysoldt-Neugebauer-Van de Walle (FNV) charge correction~\cite{freysoldt_fully_2009}, as displayed on Figure~\ref{fig:levels}\textbf{b}. 
In the non-equilibrium situation of photo-excitation, an electron is promoted from the $a$ defect level to the lowest energy $a$ level close to a conduction state. 
The corresponding electron-hole pair is made of a hole, localized at the defect position, which stems from the empty electronic level, appearing inside the bandgap after ionization of the W-center (Fig.~\ref{fig:levels}\textbf{b}).
The electron-hole interaction further pushes the transition energy inside the bandgap (Fig.~\ref{fig:levels}\textbf{c}).
As a matter of fact, light emission in W-centers, below the bandgap of silicon, results from excitonic recombination between a hole localized at the defect and an electron trapped by Coulomb interaction. 
On the contrary, in the I$_3$-I configuration, we find a degenerate $e$ level and a single $a_{1}$ level at -142~meV and -201~meV referenced to the VBM, respectively. 
These levels cannot account for a stable positive charge state of the I$_3$-I defect and thus cannot lead to a hole confined at the defect position, thus ruling out the I$_3$-I configuration as the W-center structure. 

The symmetry of the excited state for the I$_3$-V configuration has $A$ character and the ground state of the neutral defect is a closed shell $A$ state. 
Thus, the optical transition dipole moment of the I$_3$-V defect belongs to $A$ symmetry and lies in the $\left<111\right>$ direction. 
In our experiments with a detection along the $\left<001\right>$ axis, we detect the emission from the projected dipole in the (001) plane, i.e. two possible orientations either [110] or [1$\bar{1}$0] (Fig.~\ref{fig:polar}), consistently with this theoretical prediction. 
Furthermore, the extrapolation of the ZPL energy for this defect using self-consistent field method~\cite{gali_theory_2009} for two different functionals provides energies close to the experimental value of 1.018 eV for the ZPL of the W-center in silicon (see SI for detailed calculations).\\

\noindent
\textbf{Identification of the atomic configuration of the W-center in silicon.}
By considering the DFT calculations, we conclude that only the I$_3$-V configuration of the tri-interstitial defect in silicon can account for the experimental results related to W-centers in silicon including single-defect observations. 
Indeed, this atomic structure exhibits localized vibrations with the correct energy corresponding to the LVM observed in the phonon sideband emission of the W-center, and it produces a below bandgap optical emission associated with a single dipole and with a ZPL energy close to that of W-line with $A\rightarrow A$ character.


\section*{Conclusion}

In this paper, we revisit the physics of one of the most common color centers in silicon, the W-center, in light of single-defect spectroscopy and advanced DFT calculations. 
By using quantum optics tools, we demonstrate the optical isolation of single W-defects embedded in a SOI sample. 
Polarization analysis of their single-photon emission, combined with theoretical modeling, evidences that the W-center possesses a single emission dipole aligned onto its trigonal symmetry axis $\langle 111 \rangle$. 
Investigating the photophysics of these individual emitters reveals new properties associated to this important defect that are inaccessible by ensemble measurements, such as the single-photon emission statistics or the influence of the local environment on the defect photoluminescence. 
The microscopic structure of this tri-interstial complex is identified thanks to advanced DFT calculations. 
We show that the W-defect photoluminescence results from a purely excitonic mechanism leading to a below-bandgap optical emission besides the absence of electronic energy levels inside the bandgap of silicon. 
These results could enable to address new issues in material science, such as the dynamics of clustering and dissociation of interstitial aggregates in silicon~\cite{libertino_formation_2001}, or the early-stage detection of the degradation of Si-based devices in environments with strong radiations such as particle-physics detectors~\cite{macevoy_defect_1997}.

The observation of individual W-centers directly incorporated inside SOI wafers also paves the way for exploration of quantum technologies based on this intrinsic defect in silicon, specifically integrated quantum photonics and quantum communications. 
They provide a linearly-polarized single-photon emission in the near-infrared range with the high ratio of $\simeq$ 40\% directly included in the ZPL at 1.218~$\mu$m. 
 The weak ZPL wavelength dispersion between centers suggests a lower sensitivity to local strain and electrostatic variations compared to single carbon-related defects previously isolated in silicon ~\cite{redjem_single_2020, durand_broad_2021}. 
 Such property could ease achieving indistinguishability between the single photons emitted by two separated defects in future works~\cite{awschalom_quantum_2018}. 
 Photon autocorrelation measurements have also revealed the presence of a metastable level in the structure of this center. 
 Given that the ZPL corresponds to an optical emission between spin-singlets~\cite{davies_1018_1987,chartrand_highly_2018}, this metastable level could be associated with a spin triplet populated by intersystem-crossing, which invites to seek for a potential optically-detectable magnetic resonance as for the G-center in silicon~\cite{lee_optical_1982}. 
In the future, this intrinsic defect could also be created on-demand by the traditional methods of Si-implantation or irradiation, and also by the recently-developed method of laser writing~\cite{chen_laser_2017}, since it does not require the incorporation of other external elements. 
This last method could be of utmost importance since implantation steps are known to introduce Frenkel pairs in excess~\cite{ziegler_ion_1988}, which induce non-radiative recombination channels and absorption losses in the semiconductor. 
This could open the way to the controllable integration in high-Q optical microcavities and to the fabrication of efficient integrated sources of indistinguishable near-infrared single photons in SOI wafers.

Our approach combining single-defect spectroscopy and advanced first-principle calculations could be extended to explore many other color centers in silicon~\cite{davies_optical_1989,pichler_intrinsic_2004}. 
The so-called C-center, a carbon-oxygen complex giving rise to a ZPL emission at 0.790 eV (1.571 $\mu$m)~\cite{chartrand_highly_2018} or the X-center, a tetragonal cluster of four self-interstitials associated with a ZPL at 1.041 eV (1.19 $\mu$m)~\cite{hayama_photoluminescence_2004,santos_insights_2016} could be potential candidates among the hundred of fluorescent defects referenced in the broad literature on color centers in silicon~\cite{davies_optical_1989,pichler_intrinsic_2004}. 
Single-defect microspectroscopy could also lead to the discovery of new defects in silicon, as recently demonstrated for several families of single near-infrared emitters~\cite{durand_broad_2021}.

\section*{Methods}


\textbf{Sample.}
The investigated sample consists of a $^{28}$Si epilayer grown on a commercial SOI wafer, whose top Si-layer was previously thinned down to 4 nm by thermal oxidation, followed by wet hydrofluoric acid chemical etching. 
The growth of $^{28}$Si by chemical vapor deposition is described in Ref. \cite{mazzocchi_99992_2019}. 
The resulting stack is made of a 56-nm thick layer of $^{28}$Si and a 4-nm thick layer of natural Si, that are separated from the substrate by a 145-nm thick layer made of natural silicon oxide.   
To intentionally create interstitials, a square of $100\times100$ $\mu$m$^2$ has been implanted with 65-keV Si ions at a fluence of $3\times 10^{12}$ cm$^{-2}$. 
The sample underwent a flash annealing at {1000$\degree$C} during {$20\,$s} under N$_2$ atmosphere.
The areal density of single W-centers outside the implanted zone is typically of a dozen defects per $100 \times 100\, \mu$m$^2$. \\

\textbf{Experimental set-up.}
The SOI sample was measured in a scanning confocal microscope built in a closed-cycle He cryostat. 
The optical excitation was implemented above the silicon bandgap, with a CW 532-nm laser focused onto the sample via a microscope objective whose numerical aperture is NA$\,=0.85$. 
The emitted photoluminescence (PL) was collected by the same objective and sent to fiber-coupled single-photon detectors with a detection efficiency of 10\%. 
More details on the optical setup equipment can be found in \cite{redjem_single_2020}. \\


\textbf{Calculation method.}
We carried out DFT calculations using HSE06 functional~\cite{krukau_influence_2006} as implemented in the VASP plane wave based code~\cite{paier_screened_2006}. The core electrons were treated in the PAW formalism~\cite{blochl_projector_1994}. The geometric model of the defect is created in a 512-atom silicon supercell and relaxed with DFT. The large supercell allowed for using a single $\Gamma$-point sampling of the Brillouin zone. The calculated formation energies and transition levels have uncertainty about $10~\mathrm{meV}$. In the vibrational calculations, the normal modes and the phonon frequencies were obtained by using density functional perturbation theory with PBE functional~\cite{perdew_generalized_1996}. We measure the localization of the vibrational modes by calculating the inverse participation ratio (IPR) defined in Ref.~\cite{alkauskas_first-principles_2014}. The charge correction method of Freysoldt, Neugebauer and Van de Walle (FNV) was applied for the charged supercells~\cite{freysoldt_fully_2009}.

\section*{Acknowledgments}

This work was supported by the French National Research Agency (ANR) through the projects ULYSSES (No. ANR-15-CE24-0027-01), OCTOPUS (No. ANR-18- CE47-0013-01) and QUASSIC (ANR-18-ERC2-0005-01), the Occitanie region through the SITEQ contract, the German Research Foundation (DFG) through the ULYSSES project (PE 2508/1-1) and the European Union's Horizon 2020 program through the FET-OPEN project NARCISO (No. 828890) and the \mbox{ASTERIQS} project (Grant No.~820394), the National Research Development and Innovation Office of Hungary within the Quantum Technology National Excellence Program (Project Contract No.\ 2017-1.2.1-NKP-2017-00001), the National Excellence Program (Project No.~KKP129866) and the Quantum Information National Laboratory sponsored via the Ministry of Innovation and Technology of Hungary. The authors thank the Nanotecmat platform of the IM2NP institute, and Louis Hutin, Beno\^{i}t Bertrand et Shay Reboh (CEA Leti) for their contributions to the preparation of the $^{28}$SOI substrates. A. Durand acknowledges support from the French DGA.

\bibliography{W_biblio}

\end{document}